\documentclass[twocolumn,showpacs,preprintnumbers,amsmath,amssymb]{revtex4}

\usepackage{graphicx}
\usepackage{dcolumn}
\usepackage{bm}
\begin{document}
\title{Problem of the time and static restriction in quantum gravity}

\author{Shintaro Sawayama}
 \email{shintaro@th.phys.titech.ac.jp}
\affiliation{Department of Physics, Tokyo Institute of Technology, Oh-Okayama 1-12-1, Meguro-ku, Tokyo 152-8550, Japan \\
and \\
Sawayama Cram School of Physics, Atsuhara 328, Fuji-shi, Shizuoka-ken, 419-0201}
\begin{abstract}
The problem of the time is one of the open issues in the quantum gravity.
This problem is particular problem in the canonical quantum gravity.
Even in the loop gravity the problem of the time remains.
Our work is concerning to the problem of the time.
We can create a method that seems to solve this problem that is up-to-down method created in the previous works.
And we derive the static restriction in quantum gravity.
\end{abstract}

\maketitle

\section{Introduction}\label{sec1}
In the canonical quantum gravity \cite{De} there remains several open issues, that is problem of the norm and the problem of quantization the inhomogeneous spacetime and problem of the diffeomorphism. 
And the main difficulty of the quantum gravity comes from the basic equation i.e. Wheeler-DeWitt equation.
The problem of the time is one of the open issues of the canonical quantum gravity in long term period.

The problem of the time \cite{IK}\cite{SS} appeared from the formulation of canonical quantum gravity by DeWitt.
Even if we treat the loop gravity, this problem remains.
We explain what the problem of the time is in short words.\\
Even if we could solve the Wheeler-DeWitt equation, we can not know the time evolution of the states.
Solution of the quantum gravity state is the functional of only spacelike metrics, 
the state of functional does not depend on timelike metrics i.e the shift and the lapse.

The problem of the time is treated by other methods.
The one method to solve the problem of the time is to couple the other scalar field to the Hamiltonian constraint \cite{Thi}. And this paper cited many papers.
However, these method are seemed to be able to only apply to the cosmological models.
And there is problems of the gauge dependence by these methods.
We should the problem of the gauge dependence of the problem of the time.
If we coupled other field to the Wheeler-DeWitt equation, the dependence of the coordinate choice 
occur in the term of the field.
In this paper we treat full quantum gravity without fields without gauge fixing i.e our method does not depend on coordinate choice.
We create the method which is constructed by only the general relativity, that we call up-to-down method \cite{Sa}.
Once we add an another dimension as external time and we decomposes 4+1 for the direction of an external time 
and we obtain additional constraint equation and we project it usual 3+1 universes, we obtain the equation of problem of the time.
This is the explanation of what we call the up-to-down method.
In the previous works we use this method to solve the Wheeler-DeWitt equation.

In section \ref{sec2} we introduce what we call up-to-down method and we derive a main theorem of problem of the time and static restriction.
In section \ref{sec3} we comment on the problem of the time.
In section \ref{sec4} we conclude and discuss our result.
\section{Up-to-down method}\label{sec2}
In this section we rewrite what we call the up-to-down method, and some sentence is same to previous paper.
We can explain the up-to-down method shortly.
Once we add a external dimension as external time and we construct artificial functional space which
 satisfy the enlarged constraints.
Then we assume the enlarged state is constructed by the usual four dimensional quantum state.
Because the usual four dimensional quantum state satisfy the Hamiltonian and the diffeomorphism constraints,
we can obtain one additional constraint with time as one parameter from the enlarged constraints. 
We use the up-to-down method in a different way from previous papers.
We should write classical correspondence of the up-to-down method.
First we embed the usual 4-dimensional metric to the artificial 5-dimensional metric.
However, we assume the 4-dimensional Einstein equation is realized in the 5-dimensional Einstein equation.
We show the recovery of the 4-dimensional gravity is the same as the up-to-down method.
By the up-to-down method, we can obtain an additional constraint in the quantum revel which 
seems to relating to the problem of the time (Theorem 1).
However, the additional constraint of the up-to-down method is same as recovery of 4-dimensional Einstein 
gravity in classical revel. 
We show it in Lemma 1.

We start by introducing the additional dimension which is an external euclidean time with positive signature, 
and thus create an artificial enlarged functional space corresponding to this external time.
We write such external dimension as $s$.
The action may be written as
\begin{eqnarray}
S=\int _{M\times s}{}^{(5)}RdMds.
\end{eqnarray}
Where ${}^{(5)}R$ is the 5-dimensional Ricci scalar, 
built from the usual 4-dimensional metric and external time components. 
Rewriting the action by a 4+1 slicing of the 5-dimensional spacetime with lapse functional given by the $s$ direction, 
we obtain the 4+1 Hamiltonian constraint and the diffeomorphism constraints as,
\begin{eqnarray}
\hat{H}_S\equiv \hat{R}-\hat{K}^2+\hat{K}^{ab}\hat{K}_{ab} \\
\hat{H}_V^a\equiv \hat{\nabla} _b(\hat{K}^{ab}-\hat{K}\hat{g}^{ab}),
\end{eqnarray}
where a hat means 4-dimensional, 
e.g. the $\hat{K}_{ab}$ is extrinsic curvature defined by $\hat{\nabla}_a s_b$ and $\hat{K}$ is its trace, 
while $\hat{R}$ is the 4-dimensional Ricci scalar, 
and $\hat{\nabla} _a$ is the 4-dimensional covariant derivative. \\
 \\
{\it Definition 1. } The artificial enlarged functional space is defined by 
$\hat{H}_S|\Psi^{5} (g)\rangle =\hat{H}_V^a|\Psi^{5} (g)\rangle =0$, 
where $g$ is the 4-dimensional spacetime metrics 
$g_{\mu\nu}$ with ($\mu =0,\cdots ,3$).
We write this functional space as ${\cal H}_5$.
 \\ 
 
Here, the definition of the canonical momentum $P$ is different from the usual one. 
Note in fact that the above state in ${\cal H}_5$ is not the usual 5-dimensional quantum gravity state, because the 4+1 slicing is along the $s$ direction.
This fact is the reason why we call this Hilbert space as artificial functional space. 
We use the problem of the time inversely at 4+1 slicing.
By the result we can ignore the external dimension $s$.
The all components of the $s$ direction vanish by 4+1 decomposition.
It is not defined by $\partial {\cal L}/(\partial dg/dt)$ but by $\partial {\cal L}/(\partial dg/ds)$, where 
${\cal L}$ is the 5-dimensional Lagrangian. 

In addition, we impose that 4-dimensional quantum gravity must be recovered from the above 5-dimensional action.
The 3+1 Hamiltonian constraint and diffeomorphism constraint are,  
\begin{eqnarray}
H_S\equiv {\cal R}+K^2-K^{ab}K_{ab} \\
H_V^a\equiv D_b(K^{ab}-Kq^{ab}).
\end{eqnarray}
Here $K_{ab}$ is the usual extrinsic curvature defined by $D_at_b$ 
and $K$ is its trace, while ${\cal R}$ is the 3-dimensional Ricci scalar, 
and $D_a$ is the 3-dimensional covariant derivative.
Then we can define a subset of the auxiliary Hilbert space on which the wave functional satisfies the usual 4-dimensional constraints. 
In order to relate the 4 and 5 dimensional spaces we should define projections.
\\ \\
{\it Definition 2.} The subset of ${\cal H}_5$ in which the five dimensional quantum state satisfies the
extra constraints $H_S\Pi^1|\Psi ^5(g)\rangle=H_V^a\Pi^1|\Psi ^5(g)\rangle =0$
is called ${\cal H}_{5lim}$, where $\Pi^1$ is the projection
defined by
\begin{eqnarray}
\Pi^1 :{\cal H}_5 \to {\cal F}_4 \ \ \ 
\{ \Pi^1 |\Psi^5(g)\rangle=|\Psi^5(g_{0\mu}={\rm const})\rangle \} ,
\end{eqnarray}
where ${\cal F}_4$ is a functional space.
And ${\cal H}_4$ is the usual four dimensional state with the restriction that 
$H_S|\Psi^4(q)\rangle=H_V^a|\Psi ^4(q)\rangle=0$.
Here $q$ stands for the 3-dimensional metric $q_{ij}(i=1,\cdots ,3)$, and
$\Pi^2$ is defined by
\begin{eqnarray}
\Pi^2:{\cal H}_{5lim} \to {\cal H}_4 .
\end{eqnarray}
\\ 

We assume the enlargement is the multiplication of the arbitrary functional $f[g_{0\mu}]$ to the usual 4-dimensional quantum gravity state
such that,
\begin{eqnarray}
|\Psi^5 (g)\rangle =\sum_if^{(i)}[g_{0 \mu}]|\Psi_{(i)} ^4(q)\rangle .
\end{eqnarray}
The $(i)$ means $i$'s state or $i$'s functional and $|\Psi^{5(4)}(g)\rangle$ state is defined later. 
The above enlargement is the only one main assumption.
Otherwise the measure of the projection is zero.
This enlargement solves the measurement problem of the projections.

We now give a more detailed definition of the artificial functional space as follows: \\ \\ 
{\it Definition 3.} The subset ${\cal H}_{5(4)}\subset {\cal H}_5$ is defined by the constraints,  
$ H_S\Pi^3|\Psi ^5(g)\rangle =H_V^a\Pi^3|\Psi^5(g)\rangle =0$, 
and we write its elements as $|\Psi ^{5(4)}(g)\rangle$.
We also define a projection $\Pi^3$ as
\begin{eqnarray}
\Pi^3 : {\cal H}_{5(4)} \to {\cal H}_{4(5)} \ \ \  \{ P^*|\Psi ^{5(4)}(g)\rangle =|\Psi ^{5(4)}(g_{0\mu}={\rm const})\rangle
=: |\Psi ^{4(5)}(q)\rangle \} , \label{f11}
\end{eqnarray} 
where ${\cal H}_4$ is a subset of ${\cal H}_{4(5)}$.
We can defien the inner product in the ${\cal H}_{4(5)}$ space like, $\langle \Psi^{4(5)}(q)^{\dagger}|\Psi^{4(5)}(q)\rangle$
\\ 

In the next step we consider recovery of 4-dimensional quantum gravity using the decomposition of the 4-dimensional Ricci scalar.
We act the (8) state to the 4+1 Hamiltonian constraint with imposing the 3+1 constraints.
The 4-dimensional Ricci scalar is decomposed as
\begin{eqnarray}
\hat{R}=H_S+n_aH_V^a-\frac{1}{2}\dot{P}.
\end{eqnarray} 
Using the Gauss's equation
\begin{eqnarray}
\hat{R}={\cal R}-K^2+K_{ab}K^{ab}+2\nabla_a\alpha ,
\end{eqnarray}
where $\alpha$ is
\begin{eqnarray}
\alpha=n^b\nabla_bn^a-n^a\nabla_bn^b ,
\end{eqnarray}
we can rewritten the $\hat{R}$ as,
\begin{eqnarray}
\hat{R}={\cal R}+K^2-K_{ab}K^{ab}+n_a\nabla_b (P^{ab}+n^an^bK) \nonumber \\
=H_S+n_a(D_b-n_cn_b\nabla^c)(P^{ab}+n^an^bK) \nonumber \\
=H_S+n_aH_V^a-n_an_b\dot{P}^{ab}+\dot{K} \nonumber \\
=H_S+n_aH_V+\dot{K}
\end{eqnarray}
The $\dot{K}$ can be written by momentum as
\begin{eqnarray}
\dot{K}=-\frac{1}{2}\dot{P}.
\end{eqnarray}
Then the modified Hamiltonian constraint for the 5-dimensional 
quantum state which contains the 4-dimensional Einstein gravity becomes,
\begin{eqnarray}
\hat{H}_S\to -m\hat{H}_S:= -\hat{K}^2+\hat{K}^{ab}\hat{K}_{ab}-\frac{1}{2}\dot{P},
\end{eqnarray}
where $m\hat{H}$ is called modified Hamiltonian constraint simplified by using 3+1 constraint equations.
There is the theoretical branch in using the Dirac constraint or Hamiltonian and diffeomorphism constraint.
The Dirac constraint creates the additional constraint which restrict the state to be static.

Finally, the simplified Hamiltonian constraint in terms of the canonical representation becomes
\begin{eqnarray}
m\hat{H}_S =(-g_{ab}g_{cd}+g_{ac}g_{bd})\hat{P}^{ab}\hat{P}^{cd}-\frac{1}{2}\dot{P},
\end{eqnarray}
The magic constant factor $-1$ for the term $g_{ab}g_{cd}$ is a consequence of the choice of dimensions for ${\cal H}_5,{\cal H}_4$.
Here $\hat{P}^{ab}$ is the canonical momentum of the 4-dimensional metric 
$g_{ab}$, that is $\hat{P}^{ab}=\hat{K}^{ab}-g^{ab}\hat{K}$. 
And as we mentioned above, this canonical momentum is defined by the external time and not by the usual time.
We does not write $\dot{q}_{ij}$ by the commutation relation of the Hamiltonian constraint and canonical momentum at this step.

{\it Theorem 1.} In this method, in ${\cal H}_4$ additional constraint $m\hat{H}_S\Pi^3=0$ appears, if the enlargement Eq.(8) is correct.
{\it Sketch of the proof}\\
\begin{eqnarray}
\hat{H}_S\Pi^3=\hat{R}\Pi^3-\hat{K}^2\Pi^3 +\hat{K}^{ab}\hat{K}_{ab}\Pi^3 \nonumber \\
\to H_S\Pi^3 +n_aH_V^a\Pi^3-\frac{1}{2}\dot{P}\Pi^3-\hat{K}^2\Pi^3 +\hat{K}^{ab}\hat{K}_{ab}\Pi^3 \nonumber \\
\to -\frac{1}{2}\dot{P}\Pi^3-\hat{K}^2\Pi^3 +\hat{K}^{ab}\hat{K}_{ab}\Pi^3 \nonumber \\
\to (-q_{ij}g_{kl}+g_{ik}g_{jl})P^{ij}P^{kl}-\frac{1}{2}\dot{P}. \label{f12}
\end{eqnarray}
\\

Now we think the up-to-down method in the classical revel.
The 4-dimensional Einstein equation in the 5-dimensional Einstein equation becomes as follows,
\begin{eqnarray}
{}^{(4)}G_{ab}=\hat{K}\hat{K}_{ab}-\hat{K}_{a}^{\ c}\hat{K}_{bc}-2\nabla_a\beta_b=0, \label{f14}
\end{eqnarray}
where,
\begin{eqnarray}
\beta ^a:=s^b\nabla _bs^a-s^a\nabla _bs^b \label{f15}
\end{eqnarray} 
If we assume that l.h.s. of Eq. (\ref{f14}) corresponds to the matter term, we can take its trace.
And this additional constraint reduces to the sum of the  4+1 Hamiltonian constraint and the diffeomorphism constraint, that is,
\begin{eqnarray}
8\pi T^a_a:=\hat{K}^2-\hat{K}_{ab}\hat{K}^{ab}
-2\nabla_a\beta^a = m\hat{H}_S-2s_a\hat{H}_V^a-\frac{1}{2}\dot{P} \approx m\hat{H}_S. \label{f16}
\end{eqnarray}
In other words, the matter term $T_a^a$ has been promoted to the operator, 
it does not produce further constraints other than 5-dimensional modified Hamiltonian constraint. 
We don't assume equation (\ref{f14}), because it is too strong, determine the four 
independent metrics by the other metric components. \\ \\ 
{\it Lemma 1.} The requirement to recover four dimensional gravity, 
$8\pi T_a^a |\Psi ^5(g)\rangle = 0$, is the same as 
$\hat{H}_S|\Psi ^5(g)\rangle = 0$.  
So $8\pi T_a^a \approx m\hat{H}_S \approx \hat{H}_S$. 
\section{Problem of the time}\label{sec3}
The theorem of the end of section II is the main result of the problem of the time.
If we write this additional constraint in terms of operators, we obtain
\begin{eqnarray}
(q_{ab}q_{cd}-q_{ac}q_{bd})\frac{\delta}{\delta q_{ab}}\frac{\delta}{\delta q_{cd}}
-i\frac{1}{2}\frac{\partial}{\partial t}\bigg(q_{ij}\frac{\delta}{\delta q^{ij}}\bigg)=0,
\end{eqnarray}
The above equation is the main result of our work.
We would like to explain this additional equation.
In the derivation of the additional constraint equation, we implicitly solve the Wheeler-DeWitt equation.
So this equation works if we solved Wheeler-DeWitt equation at once.
If we solved the Wheeler-DeWitt equation, and if we acted it to the state, we obtain the equation which determines lapse functional.
We can think about commutation relation between the additional constraint and the Hamiltonian constraint and diffeomorphism constraint.
Off course these constraint equation does not commute.
However, the fact that the commutation relation does not hold is correct.
Because, there are following equation as
\begin{eqnarray}
\frac{d}{dt}=q_{ij,0}\frac{\delta}{\delta q_{ij}}.
\end{eqnarray}
In the commutation relation there appear an-known function $q_{ij,0}$.
So the commutation relation is closed by additional parameters.

If we treated the static spacetime, the above constraint equation takes special limit which we call static restriction, as
\begin{eqnarray}
(q_{ab}q_{cd}-q_{ac}q_{bd})\frac{\delta}{\delta q_{ab}}\frac{\delta}{\delta q_{cd}}=0.
\end{eqnarray}
The static restriction is used in the previous paper \cite{Sa2}.

Because the static restriction is the special limit of the problem of the time,
the static restriction usually does not commute with Hamiltonian and momentum constraint.
And in this case the Wheeler-DeWitt is simplified by the static restriction.
However, in some mini-super space model, the static restriction and the Hamiltonian constraint do commute.

Moreover, we can transform the additional equation as follows,
\begin{eqnarray}
t=2\int \bigg( (q_{ab}q_{cd}-q_{ac}q_{bd})P^{ab}P^{cd}\bigg)^{-1}dP\nonumber \\
=-\sum_{(i,j)\not= (k,l)}(q_{ij}P^{ij})^{-1}\ln q_{kl}P^{kl} \nonumber \\
=-\sum_{(i,j)\not= (k,l)}\bigg( -iq_{ij}\frac{\delta}{\delta q_{ij}}\bigg)^{-1}\ln \bigg( -iq_{kl}\frac{\delta}{\delta q_{kl}}\bigg) .
\end{eqnarray}
Looking the above equation, we know left hand side is only scalar, and so right hand side should be scalar.
Because above equation is operator, if the right hand side act on the state several operator become eigenvalue.
And so the right hand side is integration, this term become scalar.
It is same thing to consider commutation relation of additional constraint is to consider the commutation relation of the above equation.
The above equation may be solve the problem of the time gauge invariant way.
Note that the above equation does not comment the quantum time.
We treat the time classically.
So we can know the time by operator. 

For example we treat the following mini-superspace as
\begin{eqnarray}
g_{ab}:= \begin{pmatrix}
b & 0 & 0 & 0 \\
0 & a & 0 & 0 \\
0 & 0 & a & 0 \\
0 & 0 & 0 & a
\end{pmatrix},
\end{eqnarray}
where $b$ is $tt$ component and $a$ is $x_ix_i$ component i.e. we treat the Friedmann universe.
When the additional constraint becomes
\begin{eqnarray}
m\hat{H}_SP^*=6a^2\frac{\partial^2}{\partial a^2}
-i\frac{3}{2}\frac{\partial}{\partial t}\bigg( a\frac{\partial}{\partial a}\bigg) \nonumber \\ 
=6\frac{\partial^2}{\partial \eta^2}-
i\frac{3}{2}\frac{\partial}{\partial t}\bigg( \frac{\partial}{\partial \eta}\bigg)=0 ,
\end{eqnarray}
where $\eta$ is determined by $\eta =\log a$.

Because the usual Hamiltonian constraint is 
\begin{eqnarray}
H_S
=\frac{9}{2}a^2\frac{\partial}{\partial a^2}+\Lambda 
=\frac{9}{2}\frac{\partial}{\partial \eta^2}+\Lambda =0,
\end{eqnarray}
the state becomes
\begin{eqnarray}
|\Psi^4(\eta )\rangle =\exp (i\frac{\sqrt{2}}{3}\Lambda^{1/2} \eta (t)) .
\end{eqnarray}
where $\Lambda$ is the cosmological constant. Here, we use negative cosmological constant for example.
Using the Eq.(23) we obtain
\begin{eqnarray}
-\frac{4}{3}\Lambda+\frac{i}{3}\Lambda\dot{\eta}(t)=0.
\end{eqnarray}
From this equation we obtain
\begin{eqnarray}
\dot{\eta}(t)=4i,
\end{eqnarray}
or
\begin{eqnarray}
\eta (t)=4it.
\end{eqnarray}
Because of the above equation we obtain
\begin{eqnarray}
a(t)=\exp (4it).
\end{eqnarray}
This is the main result of the example.
The important point is $a(t)$ is given by exponential function which is consistent with inflation model.
If we use a positive cosmological constant, the form of $a$ does not change.

\section{Conclusion and discussions}\label{sec5}
We can derive equation of the time in the quantum gravity.
The independence of the time of the state can be known from the Eq.(21) by acting the quantum gravity state,
 if the Wheeler-DeWitt equation can be solved.
The Eq.(21) is the main result of our work.
Because it determined the how the spacelike metric depend on time,
this equation is suitable to quantum gravity whose spacelike metrics depend on time.
Although this equation does not solve the problem of the time,
we can derive a one additional equation relating to the time.
By the up-to-down method we can obtain the equation of the time in the gauge invariant way.

As a example we treat the Friedman universe.
Then we know how the spacelike metric $a$ depend on time $t$.
The spacelike metric $a$ is the form of exponential function.
This result coincides to the inflation models.
The example of the Friedman universe model is one of the main result of our work.

We comment on the method i.e. up-to-down method.
This method is only applied to quantum gravity, it is not applied to the usual quantum field theory.
In the previous papers of ours there are mistaken in the up-to-down method.
But we correct the mistaken and we find the up-to-down method is used for problem of the time.


\begin{thebibliography}{99}
\bibitem{De}
B.S.DeWitt Phys. Rev. {\bf 160} 1113 (1967)
\bibitem{IK}
C.J.Isham and K.V.Kuchar Ann. Phys. {\bf 164} 288 {\bf 164} 316 (1985)
\bibitem{Thi}
T.Thiemann astro-ph/0607380 (2006)
\bibitem{SS}
T.P.Shestakova and C.Simeone gr-qc/0409119 (2004)
\bibitem{Sa}
S.Sawayama gr-qc/0604007 (2006)
\end{thebibliography}
\end{document}